# First principles prediction of the Al-Li phase diagram

S. Liu[a], G. Esteban-Manzanares[a], J. LLorca[a,b,1]

[a] IMDEA Materials Institute, C/Eric Kandel 2, Getafe 28906 – Madrid, Spain

[b] Department of Materials Science. Polytechnic University of Madrid. E. T. S. de Ingenieros de Caminos. 28040 – Madrid, Spain

**Abstract**

The phase diagram of the Al-Li system was determined by means of first principles calculations in combination with the cluster expansion formalism and statistical mechanics. The ground state phases were determined from first principles calculations of fcc and bcc configurations in the whole compositional range while the phase transitions as a function of temperature were ascertained from the thermodynamic grand potential and the Gibbs free energies of the phases. Overall, the calculated phase diagram was in good agreement with the currently accepted experimental phase diagram but the simulations provided new insights that are important to optimize microstructure of these alloys by means of heat treatments. In particular, the structure of the potential GP zones, made up of $Al_{0.5}Li_{0.5}$ (001) monolayers embedded in Al matrix, was identified. It was found that $Al_3Li$ is a stable phase although the energy barrier for the transformation of $Al_3Li$ into AlLi is very small (a few meV) and can be overcome by thermal vibrations. Moreover, bcc AlLi was found to be formed by martensitic transformation of fcc configurations and $Al_3Li$ precipitates stand for favorable sites for the nucleation of AlLi because they contain the basic blocks of such fcc ordering. Finally, polynomial expressions of the Gibbs free energies of the different phases as a function of temperature and composition were given, so they can be used in mesoscale simulations of precipitation in Al-Li alloys.

---

[1] Corresponding Author.
Email address: javier.llorca@imdea.org (J. LLorca)

# 1. Introduction

Al-Li alloys have received great attention as structural materials for aerospace applications because each weight % of Li added to Al reduces the density by ≈ 3% and increases the elastic modulus by ≈ 6% [1]. Since 1920, three different generations of Al-Li have been developed. The industrial application of the first generation alloys was compromised by its limited ductility and toughness while anisotropic properties and low ductility and toughness in the short-transverse direction also hindered the use of the second generation [2]. Most of these problems have been overcome in the third generation of Al-Li alloys developed since the early 1990's and the current interest is how to improve the strength of these new alloys [3-4].

Strengthening of Al-Li alloys is mainly due to precipitation by the spherical δ' ($Al_3Li$) precipitates, which are coherent with the Al matrix. They can be sheared by dislocations and order hardening (due to the formation of antiphase boundaries) and modulus hardening (due to the mismatch in elastic modulus) contribute the most to increase the strength [5-6]. Thus, the hardening provided by δ' precipitates increases with the square root of the volume fraction and the precipitate diameter but cutting of the precipitates by successive dislocations leads to a reduction in the stress necessary to shear the precipitates and to planar slip [7].

Obviously, the Al-Li phase diagram has been carefully analyzed in the past from the experimental perspective to ascertain the precipitation sequence and the optimum heat treatments and the most relevant information is summarized in [8]. Nevertheless, there are still controversial issues that could not be answered. The first one is whether or not Guinier-Preston (GP) zones appear prior to the precipitation of δ' and what is their structure. Differential scanning calorimetry studies have detected possible GP zones that appear at room temperature and are metastable with respect to δ', dissolving in the temperature range 75-150ºC [9-10]. The precipitation and dissolution kinetics of these potential GP zones are similar to those found in GP zones in other Al alloys [11-14]. Nevertheless, they have not been observed by transmission electron microscopy because the strain contrast from these GP zones may be too small, particularly in the presence of strong contrast due to the presence of δ' precipitates.

The nucleation of δ' precipitates from a disordered solid solution has been studied from the experimental and theorical viewpoints. Floriano et al. [15] used small-angle X-ray scattering to conclude that precipitation of δ' in an Al-8.5 at.% Li was by nucleation and growth in the temperature range 60ºC to 180ºC while Osamura and Okuda [16] reported the formation of δ' in an Al-11.8 at.% Li by nucleation and growth at 260ºC and through a spinodal decomposition of the ordered phase at 140ºC. High-resolution transmission electron microscopy studies of Radmilovic et al. [17] of Al-Li alloys with Li content in the range 8.7 at.% to 10.7 at.% aged at 150ºC and 250ºC showed that δ' results from Li enrichment of the ordered regions and Li depletion of the disordered regions in the as-quenched spinodally decomposed alloys. These evidences are in agreement with the theoretical analyses of Khachaturyan et al. [18] using the mean-field model and of Soffa and Laughlin [19] based of free energy composition diagrams. Nevertheless, precise thermodynamic information (i.e. Gibbs free energies) is required to make quantitative estimations of the precise path for the formation of δ' as a function of Li content and temperature.

δ' precipitates are considered metastable [8]. This conclusion does not come from free energy calculations but from experimental observations: the equilibrium δ (AlLi) phase precipitates at grain boundaries and its growth depends on the coarsening and eventual disappearance of δ' precipitates. This process leads to the apparition of a precipitate-free zone, devoid of δ' precipitates, adjacent to the grain boundaries [3, 8, 20, 21]. It was initially assumed that δ formation was to a martensitic transformation from δ' [22] but this hypothesis was not in agreement with transmission electron microscopy observations which showed evidence of the heterogeneous nucleation of δ precipitates at grain boundaries and dislocations, independently of δ' [20, 23, 24]. Nevertheless, the growth of the δ precipitates was carried out at the expense of the Li in solid solution and in the δ' precipitates. This latter mechanism assumes that δ' is metastable but this result should be confirmed by free energy data.

The issues presented above (existence of GP zones, formation path and stability of δ') are not likely to be verified experimentally due to the uncertainties associated with the kinetics of phase transformations. These uncertainties also affect the current accepted version of the Al-Li phase diagram [25-26] constructed using the CalPhad (computation of phase diagrams) methodology from approximate analytical expressions of free energy of the different phases that are optimized from the available experimental and theoretical results [27]. Nevertheless, first principles calculations of phase diagrams are nowadays possible through the combination of *ab initio* simulations with statistical mechanics [28-34]. To this end, the formation enthalpies of the different phases at 0K are computed by means of density functional theory (DFT) and ground state phases at low temperature are determined from the convex hull of the formation enthalpy. Then, the cluster expansion Hamiltonians are fitted from the formation enthalpy data and the finite temperature thermodynamic properties of the different ground state phases are determined from low temperature expansions and Monte Carlo simulations within the semi-grand canonical ensemble [35]. This technique also enables the prediction of equilibrium short range order and vacancy concentration in ordered and disordered Al-Li alloy has been investigated using cluster expansion [36].

Previous attempts to determine the Al-Li phase diagram from first principles calculations [37-41] were hindered because of the lack of computational power. Only clusters including first and second-nearest neighbour pairs were included and neglecting higher order clusters makes difficult to predict accurately the thermodynamic properties of the different phases through Monte Carlo simulations.

These limitations were overcome in this investigation and the optimum set of clusters and the corresponding interaction coefficients were determined using a genetic algorithm implemented in CASM [42]. The ground states phases in the phase diagram were identified and special attention was given to determine the most likely structure of the GP zones in the Al-rich part of the phase diagram. It was found that the δ' phase is not metastable and the possible mechanisms for the of nucleation of δ phase were assessed by tracking structural transition paths. Finally, analytical expressions for the free energy of the different phases were provided, so they can be used as input in mesoscale models of precipitation [43-46].

## 2. Methodology

### 2.1 First principles calculation of the formation energies

The first step to calculate the Al-Li diagram is to determine the ground state phases at 0K from the formation energies of the different Al-Li intermetallic compounds. Al and Li exhibit face-centered cubic (fcc) and body-centered cubic (bcc) structures, respectively, in the stable configurations. Thus, symmetrically distinct fcc and bcc $Al_{1-x}Li_x$ configurations with Al and Li atoms randomly placed in fcc and bcc lattice sites were generated with up to 13 atoms per unit cell using CASM (Clusters Approach to Statistical Mechanics) code [42].

The relaxed energy of each $Al_{1-x}Li_x$ configuration was calculated by DFT using Quantum Espresso [47-48] in the ultra-soft pseudopotential mode [49]. The atomic positions, lattice parameters and angles of each configuration were fully relaxed at pressure P=0. The exchange-correlation energy was evaluated using the Perdew-Burke-Ernzerhof approach [50] with 114 Ry as the energy cut-off. The Brillouin zone was sampled using a Monkhorst-Pack grid with a density of 40 points/Å$^{-1}$.

The formation energy per atom of each $Al_{1-x}Li_x$ configuration can be expressed as:

$$E^f(Al_{1-x}Li_x) = E(Al_{1-x}Li_x) - (1-x)E(Al) - xE(Li) \qquad (1)$$

where $E(Al)$ and $E(Li)$ stand for the relaxed energies per atom of pure Al and pure Li in the corresponding configuration, respectively, while $E(Al_{1-x}Li_x)$ stands for the relaxed energy per atom of the corresponding configuration. This information will be used to set the effective cluster interaction (ECI) coefficients of the cluster expansion (CE) formalism, as it will be shown in the next section. It should be noted, however, that the fcc configurations with the lowest energy according to eq. (1) may not be the ground state phases because the reference formation energy of fcc Li is higher than the formation energy of the actual stable Li phase with bcc structure. The same reasoning applies to the bcc configurations because the reference formation energy of bcc Al is higher than the formation energy of the actual stable Al phase with fcc structure. Therefore, formation energy *vs.* composition map to determine the stable phases at the convex hull should be constructed using the formation energies of fcc Al and bcc Li as reference values.

**2.2 Cluster expansion for fcc and bcc structures**

The formation energies of different configurations for a given crystal lattice structure can be computed using the cluster expansion (CE) formalism [51-52] according to

$$E^f(\vec{\sigma}) = \sum_f V_f \prod_{i \in f} \sigma_i \qquad (2)$$

where $E^f(\vec{\sigma})$ is the formation energy of the configuration $\vec{\sigma}$ of a given crystal lattice with $N$ sites, which is expressed as a collection of discrete configurational variables, e.g. $\vec{\sigma} = (\sigma_1, \sigma_2, ... \sigma_N)$, where $\sigma_i = +1$ if the site $i$ is occupied by Al and $\sigma_i = -1$ otherwise (Li) in the case of a binary alloy. $\prod_{i \in f} \sigma_i$ is a specific set of crystal basis functions that describe the different types of atomic cluster interactions (pairs, triplets, quadruplets, etc.) in the system and $V_f$ stands for the effective cluster interaction (ECI) coefficients that are determined by fitting the predictions of eq. (2) from a certain number of formation energies of different configurations calculated from DFT simulations according to eq. (1). In practice, the CE converges rapidly using a relatively small number of clusters, leading to an extremely efficient strategy to determine the formation energies from the computational viewpoint.

In the case of the fcc Al-Li system, clusters of atoms with maximum atomic spacing of 8 Å for pair clusters, 8 Å for triplet clusters, and 6 Å for quadruplet clusters were considered to determine the ECI coefficients, leading to a total of 87 clusters. The optimum set of clusters and the corresponding ECIs were determined using a genetic algorithm implemented in CASM [42]. Additionally, a deep search of the preliminary ECI candidates with the best fitness score was used to avoid local minima during the optimization process as detailed in [53]. Bias and variance of the ECI obtained with the genetic algorithm were minimized via $k$-folds method, in which data is divided into $k$ sets (in this case $k$ = 20) and the ECI coefficients are calculated using $k$-1 sets as training data. Then, the resulting values are validated on the remaining set, and this process is repeated successively for all the possible combinations of $k$-1 sets to reach the optimum ECI coefficients. The final estimator of the accuracy was a cross-validation score based on a weighted least-squares fit to obtain better predictions for the structures with lower energies. Thus, a weight is given to each configuration based on the distance $d(\vec{\sigma}) = E^f_{hull} - E^f$ to the convex hull according to $w(\vec{\sigma}) = A \exp(d(\vec{\sigma})/k_b T) + B$, where A=5, B=1 and $k_b T = 0.01$ eV, where $k_b$ is the Boltzmann constant and $T$ the absolute temperature. The negative sign is implicit in the distance definition since all the formations energies ($E^f$) are negative and above the convex hull formation energy ($E^f_{hull}$).

In the case of the bcc Al-Li system, clusters of atoms with maximum atomic spacing of 8 Å for pair clusters, 6 Å for triplet clusters, and 6 Å for quadruplet clusters were considered, leading to a total of 67 clusters. The optimum set of clusters and the corresponding ECIs were determined as indicated above with $k = 30$, and A=19, B=1 and $k_bT = 0.01$ eV.

**2.3 Determination of the thermodynamics properties of phases**

The equilibrium thermodynamic properties of phases were determined through the combination of the CE formalism with statistical mechanics [51-53]. The link between the equilibrium thermodynamic properties phases and formation energies is the partition function Z in the semi-grand canonical ensemble, which is given by [54]

$$Z = \sum_s e^{-\beta(E_s^f - \Delta\mu x)N} \quad (3)$$

where $N$ is the number of sites in the crystal lattice, $\beta = 1/k_bT$, $\Delta\mu = \mu_{Al} - \mu_{Li}$ the difference in chemical potential between Al and Li, $x$ the atomic fraction of Li in the system and $E_s^f$ stands for the formation energy of each one of the possible states of the system.

Taking advantage that $E_s^f$ of each possible state can be accurately and efficiently computed through eq. (2), the partition function Z can be determined for a given temperature $T$ and chemical potential $\Delta\mu$ using the Metropolis Monte Carlo method [29]. The details of the strategy can be found in our previous publications [33-34]. The Monte Carlo calculations of both Al-Li systems started at 10K. A 'single-spin flip' low temperature expansion was first run at 10K over a large enough range of $\Delta\mu$ (-1 eV/atom to 1 eV/atom) to obtain the grand potential references for each phase [54]. Then, fine-grid metropolis Monte Carlo calculations were carried out with increasing and decreasing temperatures at increments of 10K over the range of 10K ≤ T ≤ 1000K at each chemical potential. Afterwards, the metropolis Monte Carlo calculations were performed with increasing and decreasing the chemical potentials at increments of 0.005 eV over the range of -1 eV ≤ $\Delta\mu$ ≤ 1 eV at each temperature. The Monte Carlo simulations were performed in periodic supercells of dimensions 13×13×13 primitive unit cells. For each value of $T$ and $\Delta\mu$, a Monte Carlo simulation included a number of equilibrating passes until the precision of the sampling properties reached 95%, followed by 1000 passes for calculating the thermodynamic averages. A pass is defined as N$_{sites}$ attempted flips, N$_{sites}$ being the number of sites in the Monte Carlo cell with variable occupations.

Once the partition function Z has been determined for a given temperature $T$ and chemical potential $\Delta\mu$, the thermodynamic grand potential $\Phi$ can be determined as [54]:

$$\beta\Phi = -\ln Z \tag{4}$$

where

$$\Phi = U - TS - \Delta\mu x \tag{5}$$

where $U$ and $S$ stand for the internal energy and configurational entropy of the system. Then, the phase boundary between adjacent phases can be determined. If the two adjacent phases share the same lattice structure (either fcc or bcc in this case), the stable phase for any combination of $T$ and $\Delta\mu$ is the one with the lowest $\Phi$, and the phase transition boundaries are directly determined by the intersection of $\Phi$ for both phases [53]. If the two adjacent phases have different lattice structures, the grand potentials are not directly comparable and phase transition boundaries are determined from the common tangent between the Gibbs free energies of both phases. The Gibbs free energy of the phases can be obtained from eq. (5) as

$$G = \Phi + \Delta\mu x \tag{6}$$

because the difference between Gibbs free energy and Helmholtz free energies can be neglected in the case of solid-state transformations at atmospheric pressure as the changes in the specific volume are quite small. It should be noted that the CE for fcc configurations is fitted using fcc Al and Li as references, and that for bcc configurations is fitted using bcc Al and Li as references. In order to obtain the Gibbs free energy with respect to fcc Al and bcc Li (see eq. (1)), the energy difference between fcc and bcc Al and between fcc and bcc Li should be added.

**2.4 Vibrational entropy contribution**

The stability of the different phases in the Al-Li phase diagram at finite temperatures depends on the contribution of vibrational excitations and vibrational entropy to the free energy. The vibrational contribution to the free energy $F_v$ can be expressed as [46]:

$$F_v = E_v - TS_v(T) \tag{7}$$

under the assumption that the volume of the crystal does not change with temperature, where $E_v = \int_0^\infty \frac{1}{2}\hbar\omega g(\omega)d\omega$ stands for the contribution of the vibrational excitations to the formation energy, $\hbar$ the reduced Planck's constant, $\omega$ the volume dependent phonon frequencies and $g(\omega)$ the phonon density of states. $S_v(T)$ is the entropy associated with lattice vibration, which is determined as a function of temperature by means of the quasi-harmonic approximation according to [46]:

$$S_v(T) = k_b \int_0^\infty \frac{\frac{\hbar\omega}{k_bT}}{\exp\left(\frac{\hbar\omega}{k_bT}\right)-1} g(\omega)d\omega - k_b \int_0^\infty g(\omega)\ln\left[1 - \exp\left(\frac{\hbar\omega}{k_bT}\right)\right]d\omega \tag{8}$$

$g(\omega)$ was determined by the supercell method, which is based on the calculation of the

forces on the atoms after perturbing slightly the atomic positions by using single point energy calculations [55]. Phonopy [56] was used to generate the perturbations in the atomic positions for each cell. 1 perturbation was generated in Al and Li using a 3×3×3 supercell, 2 perturbations in $Al_3Li$ using a 3×3×3 supercell, 2 perturbations in AlLi using a 2×2×2 supercell, 10 perturbations in $Al_2Li_3$ using a 2×2×2 supercell, 12 perturbations in $AlLi_2$ using a 2×2×2 supercell and 78 perturbations in $Al_4Li_9$ using a 2×2×1 supercell. After the perturbation procedure, the constant force matrix was calculated via single point calculations and then g(ω) was estimated.

The contribution of the vibrations to the free energy, $F_v^f$, was included according to

$$F_v^f(Al_{1-x}Li_x) = F_v(Al_{1-x}Li_x) - (1-x)F_v(Al) - xF_v(Li) \qquad (9)$$

**2.5 Determination of the structural transition paths**

Phase transformations connecting different crystal structures can be tracked through the energy change along the structural pathway between them [57-58]. This can be achieved by means of DFT calculations in which the atomic positions, lattice parameters and angles of a structure are fully relaxed at pressure P=0. The transformation pathway is thus shown and the energy change along the structural transformation pathway is obtained.

**3. Results**

**3.1 Formation energies at 0K**

The formation energies of 154 configurations enumerated from fcc structures (relative to the formation energies of fcc Al and Li) and of 489 configurations enumerated from bcc structures (relative to the energies of bcc Al and Li) are shown in Figs. 1(a) and (b), respectively. The analysis of the fully relaxed configurations showed that 18 configurations of the fcc system and 297 configurations of the bcc system changed the original crystal structure during relaxation and they are indicated by open symbols in Figs 1(a) and (b). The formation energy calculated by the CE formalism according to eq. (2) is restricted to a given crystal structure. Therefore, the configurations that changed the original crystal structure were not included to determine the ECI coefficients for fcc and bcc systems.

The final optimized ECIs set for the fcc Al-Li system includes 1 empty cluster interaction, 1 point cluster interaction, 4 pair cluster interactions, 6 triplet cluster interactions and 6 quadruplet cluster interaction, while 5 pair cluster interactions, 6 triplet cluster interactions and 11 quadruplet cluster interaction were are included in the case of the bcc system. The values of the ECI coefficients are depicted in Tables SI and Table SII in the Supplementary Material. The corresponding cross-validation scores of the least-squares fitting were 0.004 eV/atom for the fcc Al-Li system and 0.008 eV/atom for the bcc Al-Li system. The formation energies calculated by DFT and predicted by CE are compared in Fig. S1 in the Supplementary Material. They are in very good agreement, particularly for the structures near the convex hull.

The formation energies of both fcc and bcc configurations relative to fcc Al and bcc Li are shown in Fig. 1(c), which includes only the configurations that maintain the original lattice structure after relaxation. It should be noted that the formation energies of the fcc/bcc configurations that changed the lattice structure after relaxation were always above the convex hull in Fig. 1(c). Thus, the ground state phases of the Al-Li system at 0K are Al, $Al_3Li$ (δ'), AlLi (δ), $Al_2Li_3$, $AlLi_2$, $Al_4Li_9$ and Li, and the corresponding formation energies calculated by DFT are shown in Table I. The lattice parameters calculated in this study are also depicted in Table I together with the corresponding experimental or theoretical values from literature. In addition, there are three configurations (marked with arrows) that lie very close to the line segment connecting Al and $Al_3Li$.

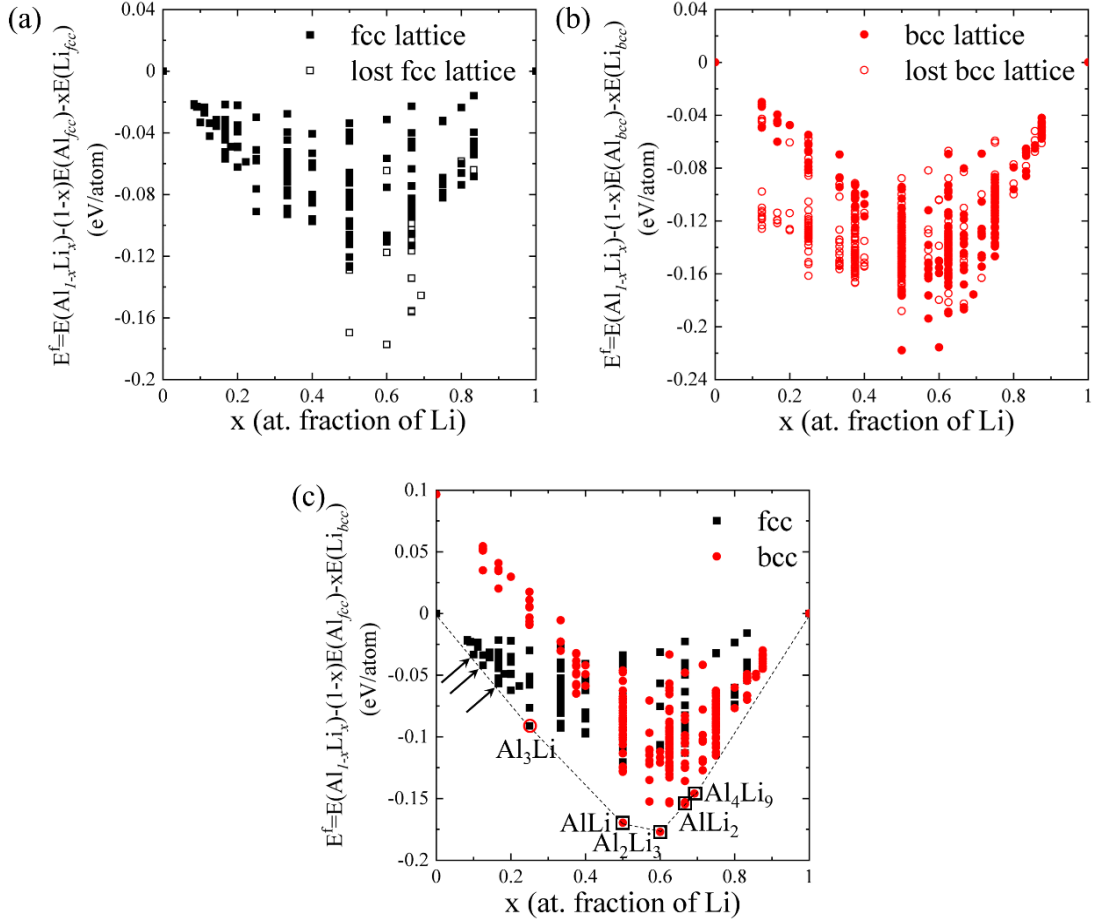

**Fig. 1.** (a) Formation energies of different configurations of fcc Al-Li system with respect to the energies of fcc Al and Li. (b) *Idem* of bcc Al-Li system with respect to the energies of bcc Al and Li. (c) *Idem* of fcc Al-Li and bcc Al-Li systems with respect to the energies of fcc Al and bcc Li. The configurations that changed the crystal structure during relaxation are marked by open symbols. The ground state phases in the convex hull are connected by dashed lines in (c). fcc $Al_3Li$ (marked

with an open circle in Fig. 1(c)) can be obtained by relaxation from a bcc configuration while bcc AlLi, $Al_2Li_3$, $AlLi_2$ and $Al_4Li_9$ (marked with open squares in Fig. 1(c)) can be obtained by relaxation from fcc configurations.

**Table I.** Formation energies (relative to fcc Al and bcc Li) and lattice parameters of Al, $Al_3Li$ (δ'), AlLi (δ), $Al_2Li_3$, $AlLi_2$, $Al_4Li_9$ and Li.

| Phases | DFT Formation energies (eV/atom) | Lattice parameters (Å) | |
|---|---|---|---|
| | | Calculated | Literature |
| fcc Al | - | a=4.02 | a=4.05 [59] |
| $Al_3Li$ (δ') | -0.0913 | a=4.00 | a=4.01 [60] |
| AlLi (δ) | -0.1696 | a=6.32 | a=6.37 [61] |
| $Al_2Li_3$ | -0.1771 | a=b=4.54 | a=b=4.44 |
| | | c=13.61 | c=14.16 [62] |
| $AlLi_2$ | -0.1547 | a=4.57 | a=4.66 |
| | | b=9.56 | b=9.77 |
| | | c=4.43 | c=4.49 [63] |
| $Al_4Li_9$ | -0.1459 | a=18.58 | a=18.92 |
| | | b=4.44 | b=4.50 |
| | | c=5.32 | c=5.42 [63] |
| bcc Li | - | a=3.43 | a=3.51 [64] |

**3.2 Vibrational contribution**

The formation energies of the ground state phases including vibrational contributions were calculated according to procedure indicated in section 2.4 and they are plotted in Fig. 2 as a function of temperature. The vibrational contribution to the formation energy is small in all cases and, moreover, is very similar for all the different phases (and, particularly, for the main strengthening phases $Al_3Li$ and AlLi). Therefore, the vibrational contribution to the free energy was not included in the phase diagram calculations.

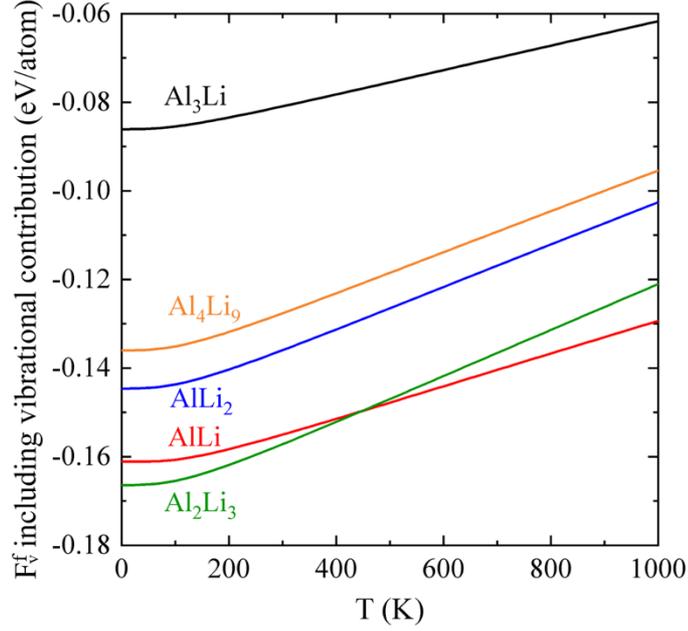

**Fig. 2.** Free energies as a function of temperature, including vibrational contributions, of $Al_3Li$ (δ'), $AlLi$ (δ), $Al_2Li_3$, $AlLi_2$, $Al_4Li_9$.

### 3.3 Martensitic transformations

A careful analysis of the configurations that lie on the convex hull in Figs. 1(a) and (b) showed that the four fcc configurations at $x=0.5$, $x=0.6$, $x=0.67$ and $x=0.69$ in the convex hull in Fig. 1(a) are relaxed into the bcc $AlLi$, $Al_2Li_3$, $AlLi_2$, $Al_4Li_9$ ground state phases, marked with open squares in Fig. 1(c). Similarly, the bcc phase in the convex hull at $x=0.25$ in Fig. 1(b) is relaxed into the fcc $Al_3Li$ ground state phase in Fig. 1(c), marked with an open circle. The bcc configuration in the convex hull at $x=0.125$ in Fig. 1(b) is above the convex hull in Fig. 1(c) and it was no longer considered.

The martensitic transformation mechanism connecting bcc and fcc phases along the Bain path is schematically shown in Fig. 3 [65]. An intermediate tetragonal lattice with lattice constants $a = \frac{1}{2}[110]_{fcc} = \sqrt{2}/2\, a_{fcc}$, $a = \frac{1}{2}[\bar{1}10]_{fcc} = \sqrt{2}/2\, a_{fcc}$ and $c = [001]_{fcc} = a_{fcc}$ is marked by red dashed lines in the fcc lattice in Fig. 3(a). The tetragonal lattice is distorted during the fcc to bcc shear transformation by expanding $a$ and contracting $c$, as shown schematically in Fig. 3(b), and the $a/c$ ratio changes from $\sqrt{2}/2$ to 1 leading to the formation of a bcc lattice. Similarly, the transformation from bcc to fcc is the inverse process that transforms a distorted bcc lattice into an intermediate tetragonal lattice in fcc and the $a/c$ ratio changes from 1 to $\sqrt{2}/2$ during the shear transformation.

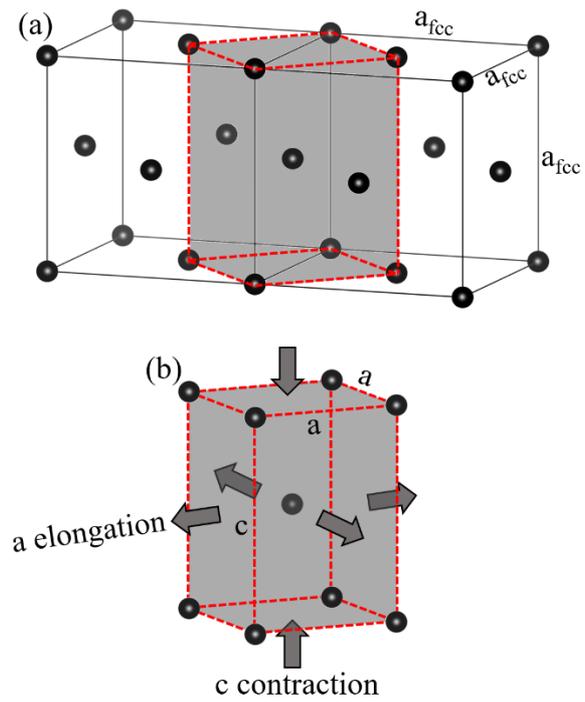

**Fig. 3.** Schematic of the martensitic transformation mechanism connecting bcc and fcc phases along the Bain path. (a) Intermediate tetragonal lattice in fcc and (b) distortion during the martensitic transformation to form the bcc lattice.

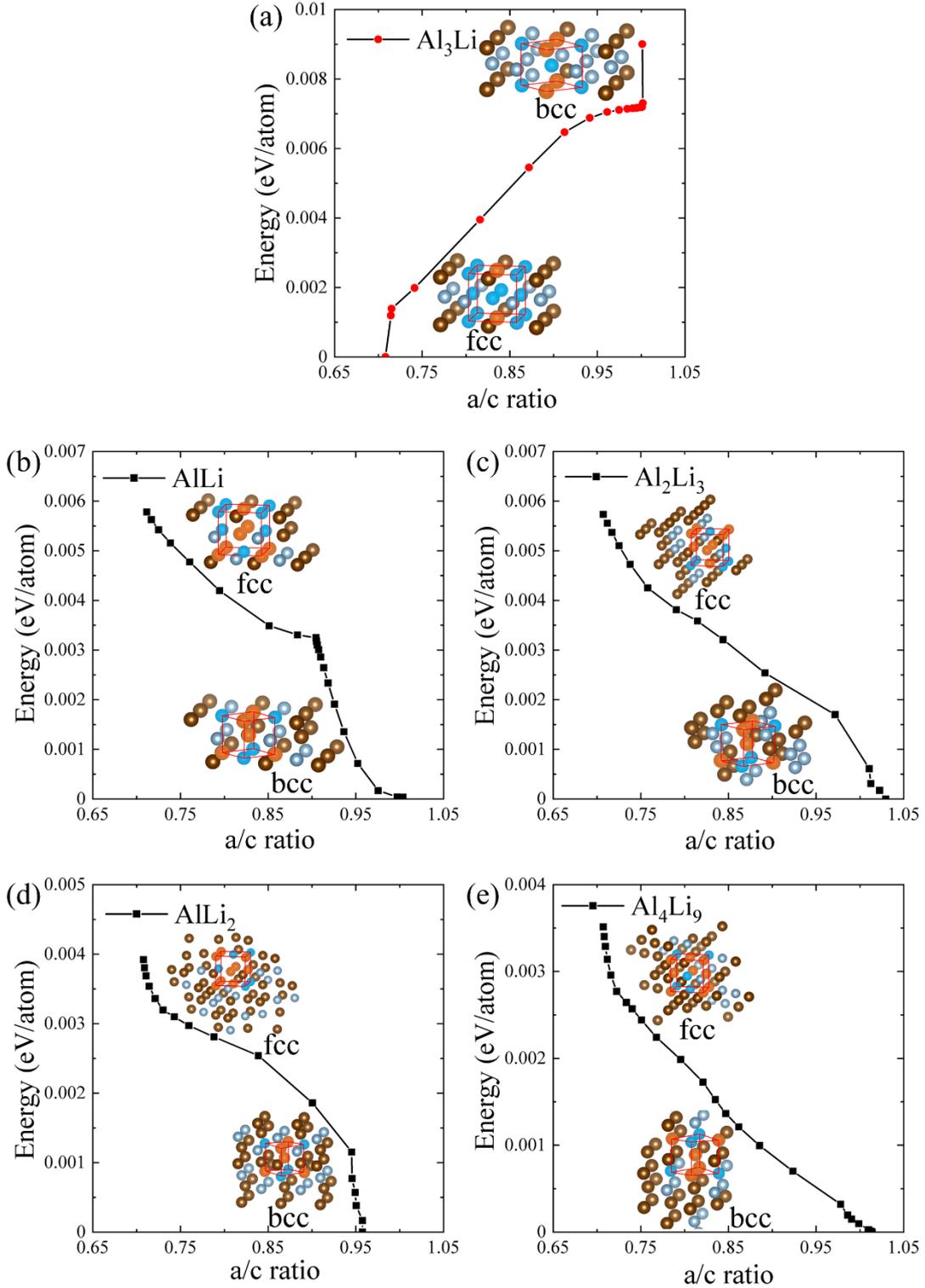

**Fig. 4.** Energy changes along Bain path of different configurations. (a) bcc to fcc transformation at $x$=0.25. (b) fcc to bcc transformation at $x$=0.5. (c) fcc to bcc transformation at $x$=0.6. (d) fcc to bcc transformation at $x$=0.67 and (e) fcc to bcc transformation at $x$=0.69. The energy is relative to the final structure of each configuration. The corresponding initial and final configurations are inserted in each figure. Al atoms are blue and Li atoms are brown. The fcc and bcc lattices are indicated in red.

The analysis of the configurations that lie on the convex hull of the fcc and bcc Al-Li system in Figs. 1(a) and (b), respectively, indicate that these martensitic transformations may take place. Therefore, the energy (per atom) changes of the configurations along the Bain path were tracked by DFT calculations for bcc to fcc transformations at $x = 0.25$ and for fcc to bcc transformations at $x = 0.5, 0.6, 0.67$ and $0.69$ and they are shown in Fig. 4. The energy paths in Fig. 4 show that there are not energy barriers that impede the transformation of the original crystal lattice into the final one. Other fcc and bcc configurations that also changed the original lattice structure during relaxation showed similar transformation paths although they did not exactly follow the Bain path.

### 3.4 Potential structure of GP zones

There are three fcc configurations at x=0.1, x=0.125 and x=0.167 in Fig. 1(c) that lie very close to the line segment connecting Al and Al$_3$Li. Their structures are plotted in Fig. 5, together with those of Al (x=0) and Al$_3$Li (x=0.25). They all have (001) planes ordering: Al contains pure Al (001) planes while the configuration at $x$=0.1 contains periodic Al$_{0.5}$Li$_{0.5}$ (001) planes separated by four layers of Al (001) planes. The configuration at x=0.125 includes periodic Al$_{0.5}$Li$_{0.5}$ (001) planes separated by three layers of Al (001) planes while periodic Al$_{0.5}$Li$_{0.5}$ (001) planes separated by three layers of Al (001) planes on one side and one Al (001) plane on the other side are found at x=0.167. Finally, Al$_3$Li is formed by alternating layers of Al$_{0.5}$Li$_{0.5}$ (001) and Al (001) planes. Structures with consecutive Al$_{0.5}$Li$_{0.5}$ (001) planes were not found in these configurations.

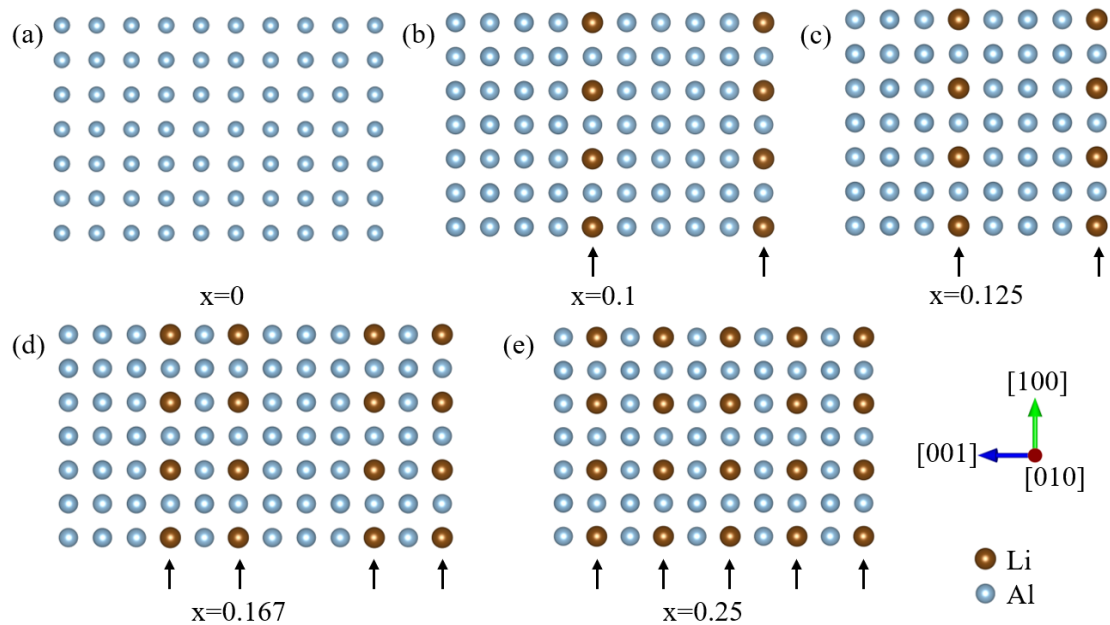

**Fig. 5.** Structure of the fcc configurations that lie on the convex hull from x=0 to x=0.25. (a) x=0 (pure Al). (b) $x$= 0.1. (c) $x$ =0.125. (d) $x$= 0.167. (e) $x$= 0.25 (Al$_3$Li). The Al$_{0.5}$Li$_{0.5}$ (001) planes are marked by arrows. Al atoms are blue and Li atoms brown.

The three configurations at $x=0.1$, $x=0.125$ and $x=0.167$ are coherent with the Al lattice and, thus, the $Al_{0.5}Li_{0.5}$ monolayer stands for a likely candidate for the GP zones on the (001) plane of Al-Li alloys. It is known that the nucleation and growth of GP zones from the supersaturated solid solution is driven by the minimization of the different energy contributions, namely chemical free energy, interface energy and elastic strain energy. The chemical free energy for the three structures with $x=0.1$, $x=0.125$ and $x=0.167$ is very similar according to Fig. 1(c) because the three of them lie parallel to the convex hull line that connects Al with $Al_3Li$. Thus, the formation energy of an $Al_{0.5}Li_{0.5}$ (001) monolayer separated by different number of Al (001) layers is weakly dependent on the number of Al (001) layers. Moreover, the interface energies of the three structures are also expected to be very close because they will come from the bonding between the $Al_{0.5}Li_{0.5}$ (001) monolayer and the surrounding Al (001) layers, which will be very similar. Finally, the elastic strain energy is controlled by the lattice mismatch. The relaxed (001) interplanar spacings of the three configurations in Fig. 5(b)~(d)) in the directions parallel ($d_\parallel$) and perpendicular ($d_\perp$) to the $Al_{0.5}Li_{0.5}$ (001) monolayers are depicted in Table II, together with the corresponding spacings of pure Al and $Al_3Li$. They are very similar in all cases and, thus, the lattice mismatch between the three potential GP configurations in Fig. 5(b)~(d)) and the Al matrix is very small, regardless of the number of Al (001) layers in between the $Al_{0.5}Li_{0.5}$ (001) monolayers. Therefore, the elastic strain energy of $Al_{0.5}Li_{0.5}$ (001) monolayers separated by different number of Al (001) layers is very low and the DFT calculations indicate that the most likely candidate for GP zones in Al-Li alloys are $Al_{0.5}Li_{0.5}$ (001) monolayers.

**Table II.** Relaxed (001) interplanar spacing in the directions parallel ($d_\parallel$) and perpendicular ($d_\perp$) to the $Al_{0.5}Li_{0.5}$ (001) monolayers.

| Configurations | $d_\parallel$ (Å) | $d_\perp$ (Å) |
|---|---|---|
| Al | 2.0078 | 2.0078 |
| x=0.1 | 1.9970 | 2.0295 |
| x=0.125 | 1.9995 | 2.0128 |
| x=0.167 | 1.9963 | 2.0152 |
| $Al_3Li$ | 2.0016 | 2.0016 |

## 3.5 Phase boundary between Al and $Al_3Li$

Al and $Al_3Li$ are the only ground-state phases in the Al-rich side of the Al-Li system according to the formation energies in Fig. 1(c). Because they share the fcc lattice, the phase boundary between them can be determined from the thermodynamic grand potentials of both phases. The grand potentials $\Phi(\Delta\mu, T)$ of Al and $Al_3Li$ were obtained as a function of temperature and chemical potential as indicated in Section 2.3 using CASM. The grand potentials $\Phi(\Delta\mu, T)$ of Al and $Al_3Li$ at T=300K are shown in Fig. 6(a), where the black curve is calculated from Al by increasing $\Delta\mu$ while the red curve is calculated from $Al_3Li$ by decreasing $\Delta \mu$. The intersection $\Phi^{Al} = \Phi^{Al_3Li}$ corresponds to chemical potential $\Delta \mu^*$ where Al and $Al_3Li$ coexist. The stable phase in the region $\Delta \mu < \Delta \mu^*$ is Al while $Al_3Li$ is stable in the region $\Delta \mu > \Delta \mu^*$. Because of the conjugated relationship between composition and chemical potential (shown in Fig. 6(b)), the composition of the coexisting Al and $Al_3Li$ phases can be determined from $\Delta \mu^*$. This procedure can be repeated at different temperatures to determine the phase boundary between Al and $Al_3Li$, which is shown in Fig. 6(c). As indicated in Section 2.3, the phase boundary can also be determined by Gibbs free energy, as schematically shown in Fig. 6(d). The dashed arrows are the tangent lines to the Gibbs free energies of Al and $Al_3Li$ at the same temperature, which stand for $\Delta \mu$. They change with composition, and when the arrows overlap, they point to the common tangent of the Gibbs free energies of Al and $Al_3Li$, namely $\Delta \mu^*$. This procedure can be repeated at different temperatures to obtain the phase boundary between Al and $Al_3Li$, which obviously coincides with the phase boundary determined from the thermodynamic grand potentials.

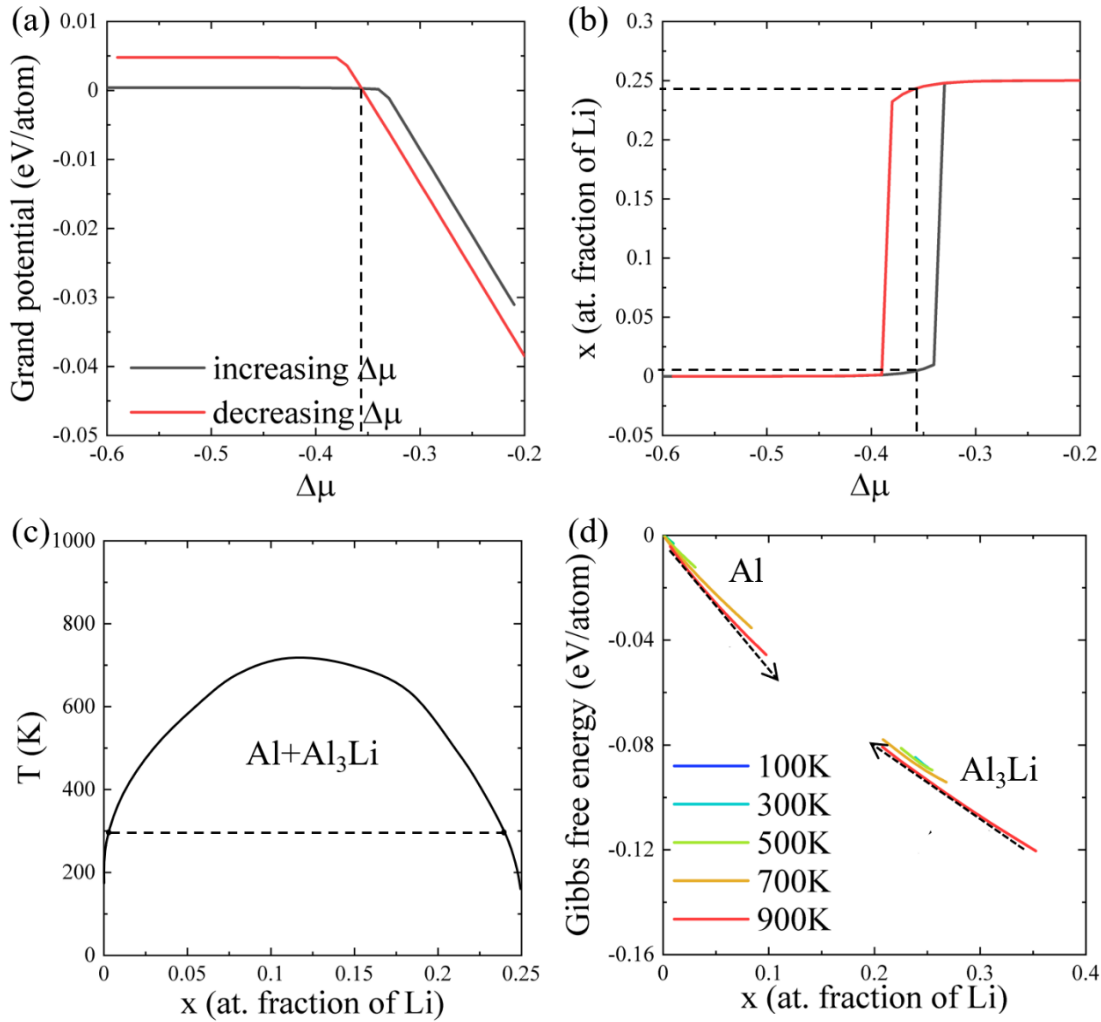

**Fig. 6** (a) Thermodynamic grand potentials $\Phi$ of Al and $Al_3Li$ as a function of the chemical potential $\Delta\mu$ at 300K. (b) Composition $x$ as a function of $\Delta\mu$ at 300K. (c) Phase transition between Al and $Al_3Li$. The dashed line corresponds to the composition at 300K according to (b). (d) Gibbs free energies of Al and $Al_3Li$ at different temperatures.

The two-phase region of Al and $Al_3Li$ exists up to T≈700K according to Fig. 6(c). Above 700K, Al and $Al_3Li$ cannot be distinguished and so the relationship between composition $x$ and $\Delta\mu$ is a continuous curve without sharp variations of $x$ at a constant $\Delta\mu$. The corresponding Gibbs free energies of Al and $Al_3Li$ also overlap above 700K, and there is no longer a common tangent between them.

### 3.6 Phase boundary between $Al_3Li$ and AlLi

As the Li content increases, the adjacent phase to $Al_3Li$ is AlLi, which has a bcc structure. The phase boundary between them can only be determined from the Gibbs free energies of both phases as a function of temperature. Their Gibbs free energies were obtained following the procedure indicated in section 2.3 and they are plotted at different temperatures in Fig. 7(a). By using the common tangent method explained above, the phase boundary between $Al_3Li$ and AlLi was determined and it is plotted in Fig. 7(b).

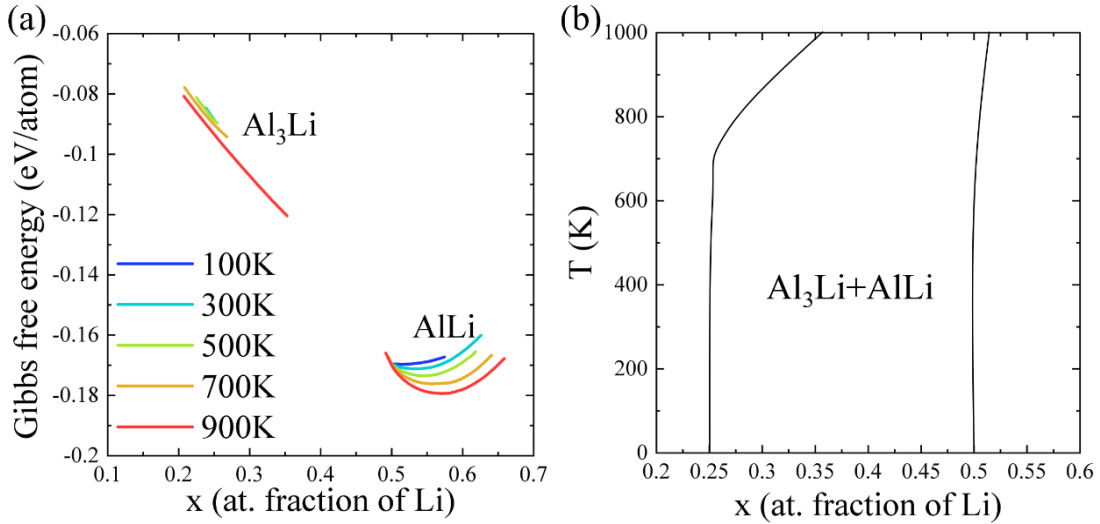

**Fig. 7** (a) Gibbs free energy of $Al_3Li$ and AlLi phases at different temperatures. (b) Phase boundary between $Al_3Li$ and AlLi.

### 3.7 Phase boundary between AlLi and $Al_2Li_3$

The phase boundary between bcc AlLi and bcc $Al_2Li_3$ was determined from the thermodynamic grand potentials, following the procedure detailed in Section 3.5. The grand potentials $\Phi(\Delta\mu, T)$ of AlLi and $Al_2Li_3$ at T=500K obtained by low temperature expansion and Monte Carlo simulations are shown in Fig. 8(a) as a function of chemical potential. The black curve is calculated from AlLi by increasing $\Delta\mu$ while the red curve is calculated from $Al_2Li_3$ by decreasing $\Delta\mu$. The conjugated relationship between composition and chemical potential is shown in Fig. 8(b), where the dashed lines indicate the composition of the coexisting AlLi and $Al_2Li_3$ phases that was determined by the condition $\Phi^{AlLi} = \Phi^{Al_2Li_3}$. The phase boundary between AlLi and $Al_2Li_3$, determined from the thermodynamic grand potentials, is plotted in Fig. 8(c). It shows that $Al_2Li_3$ is a line compound up to T≈700K, and the two-phase region also exists up to this temperature. The Gibbs free energies of AlLi and $Al_2Li_3$ are plotted in Fig. 8(d) and it shows that the Gibbs free energy of $Al_2Li_3$ is constant up to T=700K. Because the slope of the curve at both sides of x=0.6 is sharp, the common tangent between the Gibbs free energy of AlLi and $Al_2Li_3$ is always intersecting that of $Al_2Li_3$ at x=0.6 below 700K. However, the common tangent between the Gibbs free energy of AlLi and $Al_2Li_3$ intersects that of AlLi at a composition gradually approaching x=0.6 due to the large curvature of the Gibbs free energy of AlLi, and it finally intersects at x=0.6 at T=700K. Above 700K, the Gibbs free energy of AlLi is lower than that of $Al_2Li_3$, which means that AlLi is the stable phase, relative to $Al_2Li_3$, above 700K.

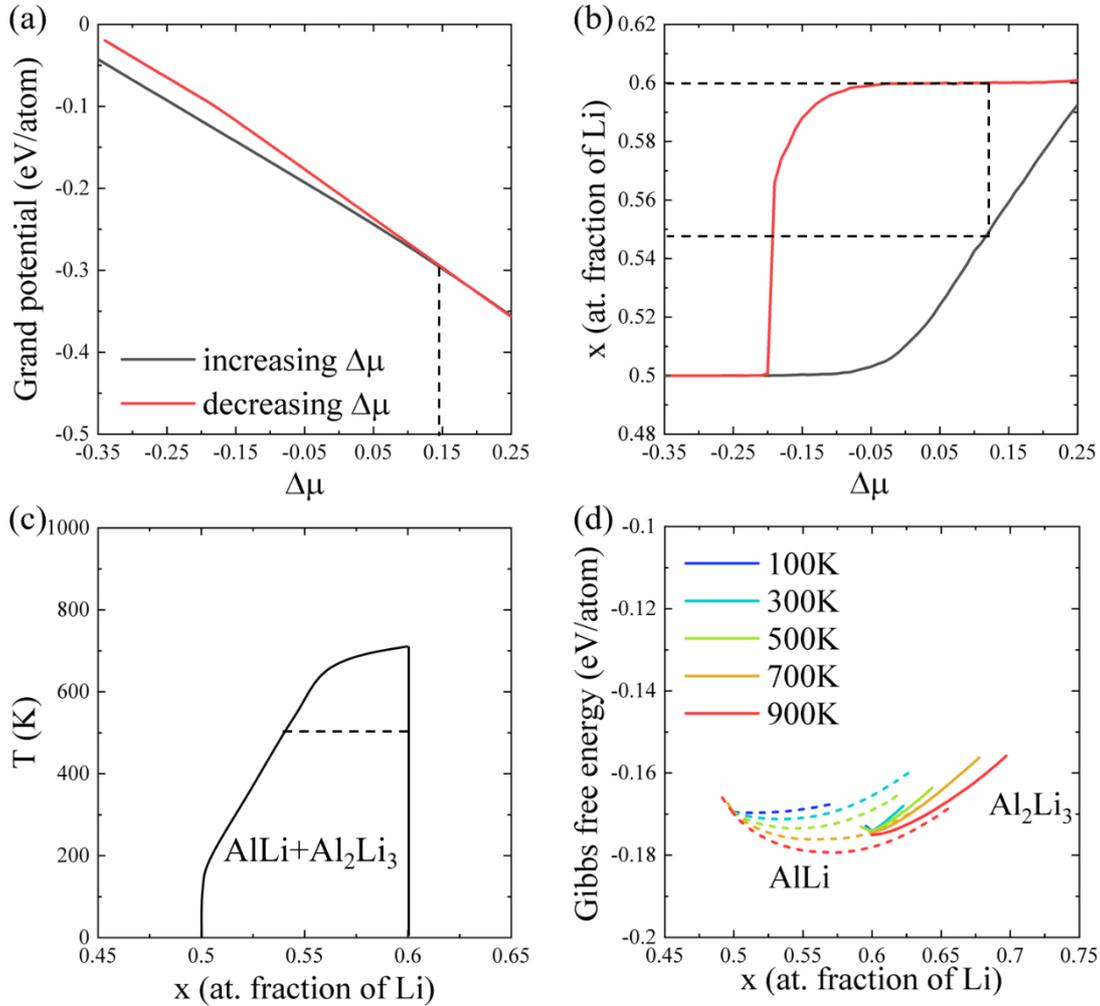

**Fig. 8** (a) Thermodynamics grand potentials $\Phi$ of AlLi and $Al_2Li_3$ as a function of the chemical potential $\Delta\mu$ at 500K. (b) Composition $x$ as a function of $\Delta\mu$ at 500K. (c) Phase transition between AlLi and Al$_2$Li$_3$. The dashed line corresponds to the composition at 500K according to (b). (d) Gibbs free energies of AlLi (dashed lines) and Al$_2$Li$_3$ (solid lines) at different temperatures.

### 3.8 Phase boundary between Al$_2$Li$_3$ and AlLi$_2$

The Gibbs free energies of Al$_2$Li$_3$ and AlLi$_2$ were computed using the same strategy and are plotted in Fig. 9(a). It is found that the Gibbs free energy of AlLi$_2$ at x=0.66 is constant up to T=500K while the Gibbs free energy of Al$_2$Li$_3$ at x=0.6 is constant up to T=700K, as mentioned above. Therefore, the common tangent between the Gibbs free energies of Al$_2$Li$_3$ and AlLi$_2$ is always intersecting at x=0.6 and x=0.66 below 300K. However, the Gibbs free energy of Al$_2$Li$_3$ is lower than that of AlLi$_2$ above 300K because the Gibbs free energy of Al$_2$Li$_3$ is decreasing more rapidly. Therefore, AlLi$_2$ is a line compound that can only exist up to 300K. Above 300K, Al$_2$Li$_3$ is stable relative to AlLi$_2$, and the corresponding phase boundary between them is plotted in Fig. 9(b).

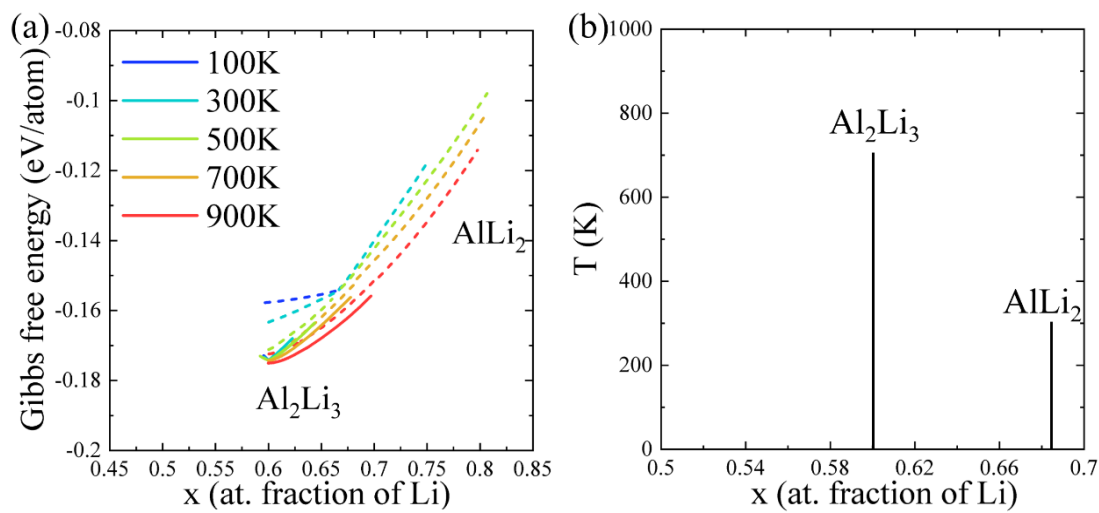

**Fig. 9** (a) Gibbs free energies of Al$_2$Li$_3$ and AlLi$_2$ at different temperatures. (b) Phase boundary between Al$_2$Li$_3$ (solid lines) and AlLi$_2$ (dashed lines).

### 3.9 Phase boundary between AlLi$_2$ and Al$_4$Li$_9$

The Gibbs free energies of AlLi$_2$ and Al$_4$Li$_9$ are plotted in Fig. 10(a) as a function of temperature. The Gibbs free energy of Al$_4$Li$_9$ at x=0.69 is constant up to T=500K. However, AlLi$_2$ and Al$_4$Li$_9$ can only coexist as line compounds up to T=300K because AlLi$_2$ is metastable relative to Al$_2$Li$_3$ above this temperature. Al$_4$Li$_9$ coexists with Al$_2$Li$_3$ between 300K and 500K, and the common tangent always intersects at x=0.6 and x=0.69. Above 500K, Al$_2$Li$_3$ is stable relative to Al$_4$Li$_9$ according to the Gibbs free energy so the corresponding phase boundary between AlLi$_2$ and Al$_4$Li$_9$ is plotted in Fig. 10(b).

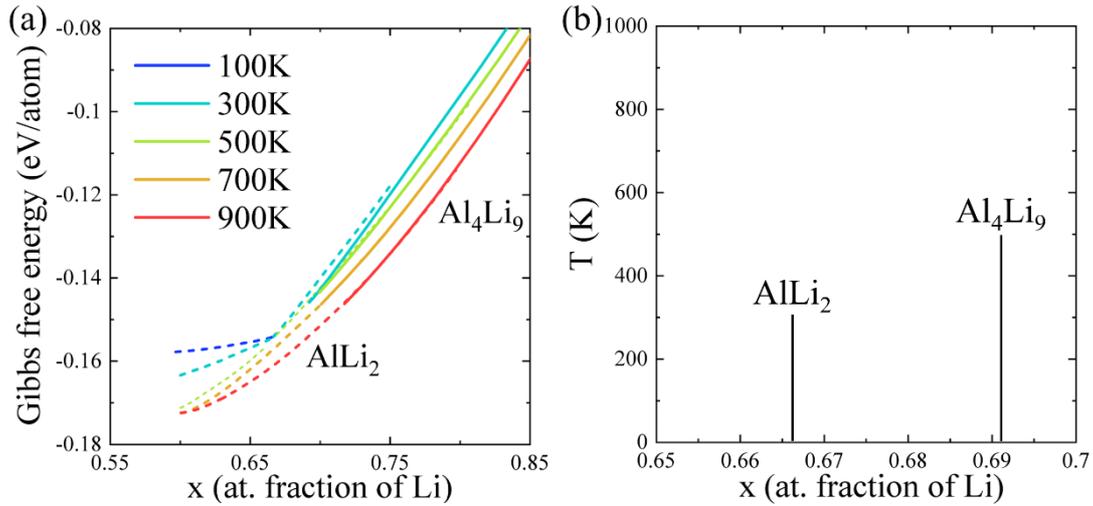

**Fig. 10** (a) Gibbs free energies of AlLi$_2$ (dashed lines) and Al$_4$Li$_9$ (solid lines) at different temperatures. (b) Phase boundary between AlLi$_2$ and Al$_4$Li$_9$.

### 3.10 Phase boundary between Al$_4$Li$_9$ and Li

Li and Al$_4$Li$_9$ have the same bcc lattice and the phase transition boundaries were directly determined by the intersection of thermodynamic grand potentials Φ for both phases, as in the case of Al and Al$_3$Li (Section 3.5) and Al$_3$Li and Al$_2$Li$_3$ (section 3.7). They are plotted in Fig. 11(a). The Gibbs free energies of Al$_4$Li$_9$ and Li are plotted in Fig. 11(b) as a function of temperature. The Gibbs free energy of Li changes with temperature, while its common tangent with the Gibbs free energy of Al$_4$Li$_9$ always intersects at x=0.69 and x=1 up to T=500K. It should be noted, however, that Li becomes liquid at ≈ 454K [25,26]. As the current investigation only deals with the solid-state transformations, the phase boundary above 500K is ignored in this region.

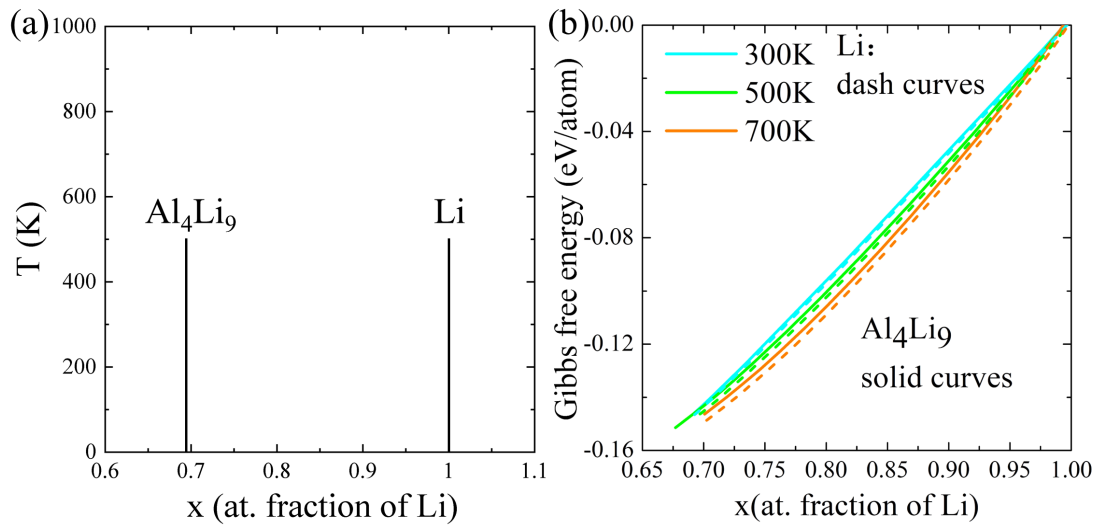

**Fig.11** (a) Phase boundary between bcc $Al_4Li_9$ and bcc Li determined from the the thermodynamic grand potential. (b) Gibbs free energies of $Al_4Li_9$ (solid lines) and Li (dashed lines) as a function of temperature.

## 4. Discussion

The analysis of the structures near to the convex hull at 0K in Fig. 1(c) indicates that Li atoms tend to be arranged as $Al_{0.5}Li_{0.5}$ (001) monolayers in the Al matrix, which can be recognized as GP zones. This is similar to the GP zones in Al-Cu alloys, which are Cu (001) monolayers in the Al matrix [11-12, 33]. The energy contributions (chemical, interface and elastic strain energy) to the nucleation of $Al_{0.5}Li_{0.5}$ (001) monolayers separated by one or several Al (001) layers are very similar and, thus, the GP zones in Al-Li alloys are essentially periodic structures composed of $Al_{0.5}Li_{0.5}$ (001) monolayers embedded in Al matrix with different spacings between them. Because the Gibbs free energy of Al and $Al_3Li$ decrease as the temperature increases (Fig. 6(d)), $Al_3Li$ with alternative $Al_{0.5}Li_{0.5}$ (001) monolayers and Al (001) monolayers will be more favorable at higher temperature. As the precipitation of $Al_3Li$ depends on the diffusion of Li atoms, it could be argued that the GP zones will be precursors for the nucleation of $Al_3Li$ because they already contain the basic blocks of $Al_3Li$, namely $Al_{0.5}Li_{0.5}$ (001) monolayers. Because the free energies of GP zones are above the common tangent of the free energies of Al and $Al_3Li$, they are always unstable with respect to $Al_3Li$. Therefore, spinodal decomposition of GP zones may occur in the absence of nucleation due to free energy changes caused by composition fluctuations [66]. These results are in good agreement with detection of a metastable phase with respect to $Al_3Li$ that appears at room temperature and dissolves in the temperature range 350K to 425K in an Al-2.5% Li alloy [9-10].

From Table I, the difference in the free energy of $Al_3Li$ and the straight line connecting Al and AlLi is very small (-6.46 meV/atom from DFT, -6.41 meV/atom from the energies calculated by the CE formalism) but these values are of the same order of the accuracy of the DFT calculations. Errors associated with the fitting of the CE could be neglected because the cross-validation scores were very small (4 meV/atom for the fcc Al-Li) and the differences between DFT and CE energies are much smaller near the convex hull because of the weighted least-squares fit strategy. If the contribution of the vibrational excitations to the formation energy, $E_v$, is included, the difference in the formation energy of $Al_3Li$ and the straight line connecting Al and AlLi at 0K decreases to -5.5 meV. Finally, the vibrational entropic contribution to the free energy of $Al_3Li$ is slightly higher than that of AlLi (Fig. 2) but the difference is also very small (around -3 meV at T = 300K). Thus, the first principles calculations indicate that $Al_3Li$ is a stable phase but the energy barrier for the transformation of $Al_3Li$ into AlLi is indeed very small.

According to the transformation paths depicted in Fig. 4, martensitic transformations occur in the Al-Li system for configurations with specific order. Special attention should be given to Fig. 4(b), in which AlLi (bcc) is formed from fcc ordering. The structures of AlLi before and after transformation are shown Fig. 12, in which the corresponding lattice vectors are marked accordingly. The (001) planes of the fcc AlLi are $Al_{0.5}Li_{0.5}$ (001) layers, equivalent to the $Al_{0.5}Li_{0.5}$ (001) monolayers in $Al_3Li$. The very small energy barrier for the transformation of $Al_3Li$ into AlLi may be overcome by thermal vibrations at high temperature, and $Al_3Li$ precipitates stand for favorable sites for the nucleation of AlLi because such fcc ordering in Fig. 12(a) is necessary for bcc AlLi (Fig. 12(b)) and $Al_3Li$ already contains the basic blocks of such fcc ordering. Thus, precipitation of AlLi can be accompanied by the coarsening and disappearance of $Al_3Li$ precipitates, as reported experimentally [3,8]. Moreover, the following orientation relationships between Al matrix and AlLi can be predicted from the lattice vectors in Fig. 12: $(001)_{Al}//(001)_{AlLi}$, $(100)_{Al}//(110)_{AlLi}$, $(010)_{Al}//(1\bar{1}0)_{AlLi}$ and from the last two orientations, another potential orientation relationship is possible: $(110)_{Al}//(100)_{AlLi}$. In addition, according to the martensitic transformation mechanism, the angle between the (111) plane of Al and the (011) plane of AlLi is just 10°. They are in good agreement with the $(110)_{Al}//(100)_{AlLi}$ and $(111)_{Al}//(011)_{AlLi}$ orientation relationships determined experimentally [61].

Finally, the shear strain associated with the fcc to bcc martensitic transformation can also be responsible for the preferential nucleation of AlLi precipitates on matrix dislocations and grain boundaries [23-24], as it was demonstrated for θ' ($Al_2Cu$) precipitates in Al-Cu alloys. The nucleation of θ' precipitates is associated with an invariant plane strain structural transformation with a large shear component [67] and the interaction between the stress fields of the precipitate and dislocation/grain boundary facilitates the heterogeneous nucleation of the precipitates on these defects [45]. This mechanism is not active, however, in the case of $Al_3Li$ precipitates - which precipitate homogeneously in the matrix [15-17] - because its formation is not associated to a shear transformation strain.

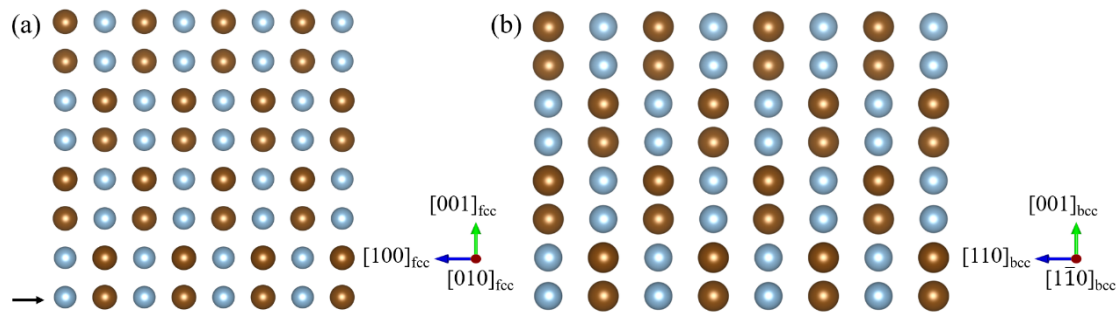

**Fig. 12** Structures of AlLi (a) before the martensitic transformation (fcc). (b) After the martensitic transformation (bcc). Al atoms are blue and Li atoms brown. Each plane in (a) -as marked by the arrow- is $Al_{0.5}Li_{0.5}$ (001) plane.

The Gibbs free energies of α-Al, $Al_3Li$ and AlLi as a function of the atomic fraction of Li $x$ and the absolute temperature $T$ in Figs. 5 and 6 have been adjusted to polynomial functions which can be found in the Table SIII of the Supplementary Material, so they can be used as input in mesoscale simulations of the precipitation in the Al-Li system [46, 68]. The other phases are line compounds in the range of temperatures of interest, and their Gibbs free energies are given by the formation energies at 0K. The currently accepted version of the Al-Li phase diagram, constructed from experimental data [25-26] and the Al-Li phase diagram calculated from first principles are compared in Fig. 14. $Al_3Li$ is considered a metastable phase in the experimental phase diagram, while it is stable in the calculated one. The calculated right border of the Al-$Al_3Li$ two-phase region is in very good agreement with the experimental one. There is a gap between the Al region and Al-$Al_3Li$ two-phase region in the experimental phase diagram that, based on the thermodynamics restrictions, cannot be strictly correct. In the experimental phase diagram, AlLi is stable in a large compositional range and the left border of AlLi region (at $x≈0.45$) is also the right border of Al-AlLi two-phase region. However, the left border of AlLi region has to be the right border of $Al_3Li$-AlLi two-phase region because $Al_3Li$ is a stable phase. Based on thermodynamic restrictions, the left border of the AlLi region will have to move in the direction of increasing Li content compared with the experimental phase boundary, which is consistent with the calculated phase diagram. The composition of the right border of AlLi region in the experimental phase diagram is higher than the calculated one, and the line compounds $Al_2Li_3$ and $Al_4Li_9$ can exist up to a higher temperature in the experimental one, which may due to the kinetic limitations associated with the experiments. $AlLi_2$ is not shown in the experimental phase diagram, which may due to the similarity in composition with $Al_2Li_3$ and $Al_4Li_9$, but it was recently reported by Puhakainen et al. [63]. It should be finally noted that the methodology to calculate the phase diagram in this paper is limited to solid-state transformations and ignores the phase boundaries related to liquid phases.

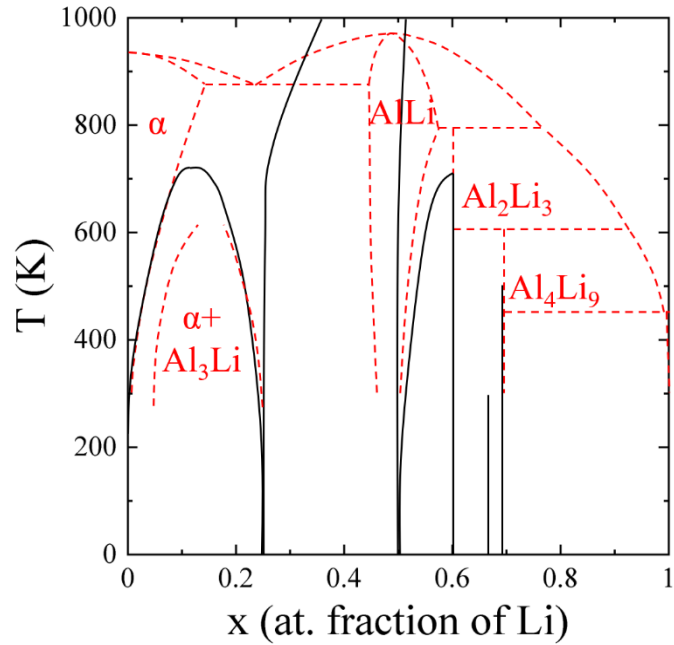

**Fig.13** Comparison between the currently accepted Al-Li phase diagram [25-26] and the calculated phase diagram. The experimental phase diagram is indicated by dashed red curves while the calculated phase diagram is delineated by solid black curves.

## 5. Conclusions

The Al-Li phase diagram was predicted from first principles calculations in combination with statistical mechanics. From the formation energies of different fcc and bcc configurations (calculated by DFT), the ground state phases and the effective cluster interaction coefficients of the cluster expansion Hamiltonian were determined. Then, the thermodynamic grand potential and Gibbs free energy of the different phases as a function of temperature were obtained from the partition function calculated by Metropolis Monte Carlo simulations. Overall, the phase diagram calculated from first principles was in good agreement with the accepted experimental Al-Li phase diagram but the analysis of the simulation results provided new insights in the actual phases and stability of this system. In particular:

• The structure of the potential GP zones was identified: they are formed by periodic structures composed of $Al_{0.5}Li_{0.5}$ (001) monolayers embedded in Al matrix with different spacings between them. Moreover, it is likely that these GP zones are precursors for the nucleation of $Al_3Li$ because they already contain the basic blocks of $Al_3Li$, namely $Al_{0.5}Li_{0.5}$ (001) monolayers.

• It was found that $Al_3Li$ is a stable phase although the energy barrier for the transformation of $Al_3Li$ into AlLi is very small (a few meV) and can be overcome by thermal vibrations. AlLi (bcc) precipitates are formed by a martensitic transformation of fcc configurations and Al$_3$Li precipitates stand for favorable sites for the nucleation of AlLi because they contain the basic blocks of such fcc ordering. This mechanism explains why the precipitation of AlLi is always accompanied by the coarsening and disappearance of Al$_3$Li precipitates. Moreover, the shear strain associated to the martensitic fcc-bcc transformation to form AlLi is responsible for its heterogeneous nucleation on dislocation and grain boundaries.

• The calculated phase diagram corrects some inconsistencies in the experimental phase diagram. Moreover, it predicts that the line compounds Al$_2$Li$_3$ and Al$_4$Li$_9$ are stable up to higher temperatures than those indicated in the experimental phase diagram and a new line compound, AlLi$_2$, is found between them.

• Finally, polynomial expressions of the Gibbs free energies of the different phases as a function of temperature and composition were obtained, so they can be used in mesoscale simulations of precipitation in Al-Li alloys.

## Acknowledgements

This investigation was supported by the European Research Council (ERC) under the European Union's Horizon 2020 research and innovation programme (Advanced Grant VIRMETAL, grant agreement No. 669141). SL acknowledges the support from the European Union's Horizon 2020 research and innovation programme through a Marie Sklodowska-Curie Individual Fellowship (Grant Agreement 893883). Computer resources and technical assistance provided by the Centro de Supercomputación y Visualización de Madrid (CeSViMa) and by the Spanish Supercomputing Network (project FI-2020-2-0044, node Calendula) are gratefully acknowledged. Finally, use of the computational resources of the Center for Nanoscale Materials, an Office of Science user facility, supported by the U.S. Department of Energy, Office of Science, Office of Basic Energy Sciences, under Contract No. DE-AC02-06CH11357, is also gratefully acknowledged.

## Conflict of interest statement

On behalf of all authors, the corresponding author states that there is no conflict of interest.

Supplementary Material for

# First principles prediction of the Al-Li phase diagram

S. Liu[a], G. Esteban-Manzanares[a], J. LLorca[a,b]

[a] IMDEA Materials Institute, C/Eric Kandel 2, Getafe 28906 – Madrid, Spain

[b] Department of Materials Science. Polytechnic University of Madrid. E. T. S. de Ingenieros de Caminos. 28040 – Madrid, Spain

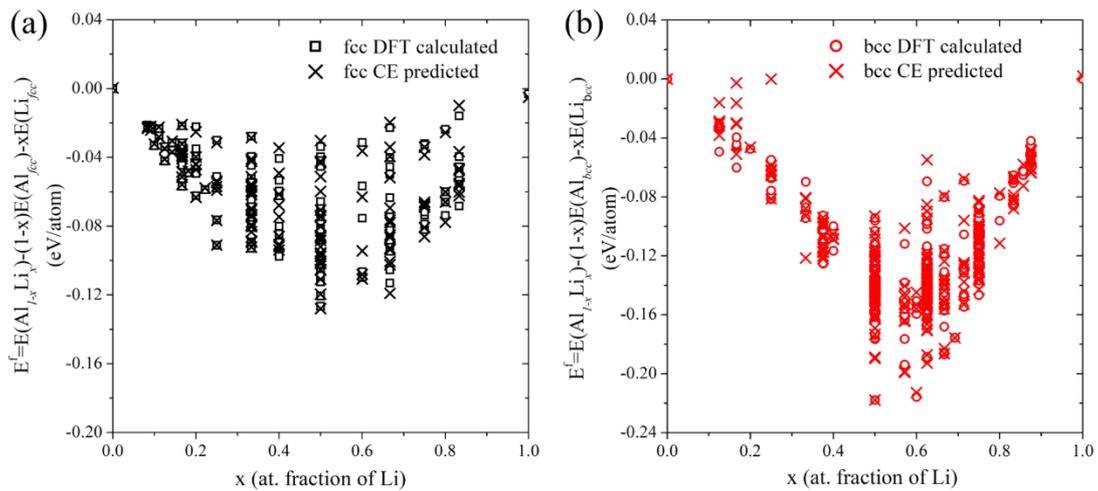

**Fig. S1.** (a) Comparison of the DFT calculated and CE predicted formation energies of different configurations of fcc Al-Li system with respect to the energies of fcc Al and Li. (b) *Idem* of bcc Al-Li system with respect to the energies of bcc Al and Li.

Table SI. Effective cluster interactions coefficients of fcc Al-Li system.

| Cluster | Sites (Fractional Coordinates of fcc primitive cell) | Max Distance (Å) | Min Distance (Å) | Multiplicity | ECI (eV/atom) |
|---|---|---|---|---|---|
| 1 |  | 0 | 0 | 1 | 0.0004 |
| 2 | 0, 0, 0 | 0 | 0 | 1 | -0.2162 |
| 3 | 0, 0, 0<br>0, 0, -1 | 2.8395 | 2.8395 | 6 | 0.1798 |
| 4 | 0, 0, 0<br>1, -1, -1 | 4.0156 | 4.0156 | 3 | -0.0660 |
| 5 | 0, 0, 0<br>0, 0, -2 | 5.6790 | 5.6790 | 6 | -0.1439 |
| 6 | 0, 0, 0<br>1, -1, -2 | 6.3493 | 6.3493 | 12 | 0.0909 |
| 7 | 0, 0, 0<br>1, -1, -1<br>0, 0, -1 | 4.0156 | 2.8395 | 12 | -0.2663 |
| 8 | 0, 0, 0<br>1, 0, -2<br>1, -1, -1 | 4.9181 | 2.8395 | 24 | 0.0694 |
| 9 | 0, 0, 0<br>0, 0, -2<br>1, 0, -2 | 5.6790 | 2.8395 | 48 | 0.2051 |
| 10 | 0, 0, 0<br>0, 0, -2<br>2, 0, -2 | 5.6790 | 5.6790 | 8 | 0.0606 |
| 11 | 0, 0, 0<br>1, -1, -2<br>2, -1, -2 | 6.3493 | 2.8395 | 24 | -0.1305 |
| 12 | 0, 0, 0<br>1, 0, -3<br>-1, 1, -2 | 7.5126 | 4.9181 | 48 | -0.1198 |
| 13 | 0, 0, 0<br>0, 0, -1<br>1, 0, -1<br>0, 1, -2 | 2.8395 | 2.8395 | 2 | 0.0743 |
| 14 | 0, 0, 0<br>1, -1, -1<br>0, 0, -1<br>1, -1, 0 | 4.0156 | 2.8395 | 3 | 0.0558 |
| 15 | 0, 0, 0 | 4.9181 | 2.8395 | 12 | 0.1127 |

| | 1, 0, -2 | | | | |
| | 0, 0, -1 | | | | |
| | 1, 0, -1 | | | | |
| 16 | 0, 0, 0<br>0, 0, -2<br>1, 0, -2<br>0, 0, -1 | 5.6790 | 2.8395 | 48 | -0.2436 |
| 17 | 0, 0, 0<br>0, 0, -2<br>1, 0, -2<br>1, 0, -1 | 5.6790 | 2.8395 | 24 | 0.1472 |
| 18 | 0, 0, 0<br>0, 0, -2<br>1, -1, -1<br>0, 0, -2 | 5.6790 | 2.8395 | 12 | 0.1510 |

**Table SII.** Effective cluster interactions coefficients of bcc Al-Li system.

| Cluster | Sites (Fractional Coordinates of bcc primitive cell) | Max Distance (Å) | Min Distance (Å) | Multiplicity | ECI (eV/atom) |
|---|---|---|---|---|---|
| 1 | 0, 0, 0<br>-1, -1, -1 | 2.9678 | 2.9678 | 4 | -0.9208 |
| 2 | 0, 0, 0<br>0, -1, -1 | 3.4268 | 3.4268 | 3 | -0.3093 |
| 3 | 0, 0, 0<br>-2, -2, -2 | 5.9354 | 5.9354 | 4 | -0.2403 |
| 4 | 0, 0, 0<br>-2, -2, -3 | 7.4686 | 7.4686 | 12 | -0.0479 |
| 5 | 0, 0, 0<br>-1, -2, -3 | 7.6626 | 7.6626 | 12 | -0.0764 |
| 6 | 0, 0, 0<br>-1, -1, -1<br>0, -1, -1 | 3.4268 | 2.9677 | 12 | 1.2777 |
| 7 | 0, 0, 0<br>-1, -2, -2<br>-1, -1, -1 | 5.6827 | 2.9677 | 24 | 0.4251 |
| 8 | 0, 0, 0<br>-1, -2, -2<br>0, -1, -2 | 5.6827 | 3.4268 | 24 | 0.0667 |
| 9 | 0, 0, 0<br>-2, -2, -2 | 5.9354 | 2.9677 | 4 | 0.4390 |
| | -1, -1, -1 | | | | |
| 10 | 0, 0, 0<br>-2, -2, -2<br>-1, -2, -2 | 5.9354 | 2.9677 | 24 | 0.1895 |
| 11 | 0, 0, 0<br>-2, -2, -2<br>-1, -1, -2 | 5.9354 | 3.4268 | 24 | 0.3154 |
| 12 | 0, 0, 0<br>-1, -2, -2<br>-1, -1, -2<br>-1, -1, -1 | 5.6827 | 2.9677 | 48 | -0.2689 |
| 13 | 0, 0, 0<br>-1, -2, -2<br>-2, -1, -2<br>-1, -2, -1 | 5.6827 | 2.9677 | 48 | -0.2468 |
| 14 | 0, 0, 0<br>-1, -2, -2<br>-1, -1, -2<br>0, 0, -1 | 5.6827 | 2.9677 | 24 | 0.2781 |
| 15 | 0, 0, 0<br>-1, -2, -2<br>0, -1, -2<br>0, -1, -1 | 5.6827 | 2.9677 | 24 | 0.0876 |
| 16 | 0, 0, 0<br>-1, -2, -2<br>-1, -1, -1<br>0, -1, -1 | 5.6827 | 2.9677 | 12 | -0.3560 |
| 17 | 0, 0, 0<br>-1, -2, -2<br>-1, -1, -2<br>0, -1, 0 | 5.6827 | 2.9677 | 12 | 0.0664 |
| 18 | 0, 0, 0<br>-2, -2, -2<br>-1, -1, -2<br>-1, -1, -1 | 5.9354 | 2.9677 | 24 | -0.4681 |
| 19 | 0, 0, 0<br>-2, -2, -2<br>-1, -2, -2<br>0, 0, -1 | 5.9354 | 2.9677 | 24 | -0.1932 |
| 20 | 0, 0, 0<br>-2, -2, -2<br>-1, -2, -2<br>-1, -1, -1 | 5.9354 | 2.9677 | 24 | -0.3857 |
| 21 | 0, 0, 0 | 5.9354 | 2.9677 | 24 | 0.2917 |
|    | -2, -2, -2 |        |        |    |        |
|    | -1, -2, -2 |        |        |    |        |
|    | -2, -1, -2 |        |        |    |        |
| 22 | 0, 0, 0  | 5.9354 | 2.9677 | 48 | 0.0783 |
|    | -2, -2, -2 |        |        |    |        |
|    | -1, -2, -2 |        |        |    |        |
|    | -1, -1, -2 |        |        |    |        |

**Table SIII.** Coefficients of the polynomial fit to the Gibbs free energies of α-Al, Al$_3$Li and AlLi as a function of the atomic fraction of Li x at different temperatures.

| | Polynomial form of the Gibbs free energy of α-Al below 700K: | | | | |
| | $G = A + B_1 \cdot x + B_2 \cdot x^2 + B_3 \cdot x^3 + B_4 \cdot x^4$ | | | | |
| T(K) | A | $B_1$ | $B_2$ | $B_3$ | $B_4$ |
|---|---|---|---|---|---|
| 0 | 4.02E-04 | | | | |
| 50 | 4.02E-04 | | | | |
| 100 | 4.01953E-4 | -0.305 | | | |
| 150 | 4.0073 | -0.33012 | 10.59752 | | |
| 200 | 3.96721E-4 | -0.3811 | 43.5048 | -16113.10158 | 2324445.32463 |
| 250 | 3.99997E-4 | -0.41797 | 38.1567 | -8926.09143 | 762358.28553 |
| 300 | 3.98077E-4 | -0.43801 | 22.96826 | -2813.70789 | 127363.99662 |
| 350 | 3.92281E-4 | -0.46404 | 17.55599 | -1437.7134 | 45485.78617 |
| 400 | 3.90139E-4 | -0.48772 | 15.43516 | -1016.52036 | 25731.65434 |
| 450 | 3.81529E-4 | -0.50385 | 10.82044 | -436.07099 | 6770.33791 |
| 500 | 3.72132E-4 | -0.52542 | 9.79033 | -332.01536 | 4411.08409 |
| 550 | 3.41465E-4 | -0.53215 | 6.92629 | -162.85024 | 1546.99823 |
| 600 | 3.05867E-4 | -0.54586 | 5.71035 | -101.44573 | 720.81492 |
| 650 | 2.60296E-4 | -0.55825 | 4.8603 | -68.84575 | 396.21234 |
| | Polynomial form of the Gibbs free energy of Al$_3$Li below 700K: | | | | |
| | $G = A + B_1 \cdot x + B_2 \cdot x^2 + B_3 \cdot x^3 + B_4 \cdot x^4$ | | | | |
| 0 | -0.0885 | | | | |
| 50 | -0.0885 | | | | |
| 100 | 0.00276 | -0.36502 | | | |

| | | | | | | |
|---|---|---|---|---|---|---|
| 150 | 0.30672 | -2.81228 | 4.92566 | | | |
| 200 | 7328.75201 | -118276.4095 | 715801.13683 | -1925319.93574 | 1941968.59061 | |
| 250 | 0.35798 | -3.26923 | 5.93325 | | | |
| 300 | 1798.39111 | 83200.02154 | 1443467.8925 | 1.11296E7 | 3.21785E7 | |
| 350 | 93.98677 | -1564.9956 | 9766.9856 | -27095.513 | 28186.73041 | |
| 400 | 42.83536 | -719.79877 | 4531.33491 | -12685.13799 | 13317.79744 | |
| 450 | 13.47723 | -230.79859 | 1477.76989 | -4212.95022 | 4505.74599 | |
| 500 | 2.38507 | -43.43123 | 291.38839 | -875.92013 | 987.78713 | |
| 550 | -0.86309 | 12.82084 | -73.59675 | 175.41798 | -146.44346 | |
| 600 | -0.76159 | 12.07027 | -75.22974 | 198.50135 | -190.65078 | |
| 650 | -0.63237 | 10.34993 | -67.6028 | 186.94078 | -189.50362 | |

Polynomial form of the Gibbs free energies of Al and $Al_3Li$ above 700K:

$$G=A+ B_1*x+ B_2*x^2+ B_3*x^3+ B_4*x^4$$

| | | | | | |
|---|---|---|---|---|---|
| 700 | 4.75788E-5 | -0.52795 | 1.93261 | -9.18505 | 16.51366 |
| 750 | -7.66638E-5 | -0.5391 | 1.76872 | -7.04811 | 11.46084 |
| 800 | -3.0529E-4 | -0.53914 | 1.39062 | -3.84314 | 4.60349 |
| 850 | -4.11442E-4 | -0.55794 | 1.45429 | -3.69966 | 3.92965 |
| 900 | -4.72526E-4 | -0.57917 | 1.5822 | -4.08066 | 4.30022 |
| 950 | -6.626729E-4 | -0.5946 | 1.6225 | -4.10975 | 4.25616 |
| 1000 | -6.62763E-4 | -0.61551 | 1.70652 | -4.28112 | 4.3795 |

Polynomial form of the Gibbs free energy of AlLi:

$$G=A+ B_1*x+ B_2*x^2+ B_3*x^3+ B_4*x^4+ B_5*x^5+ B_6*x^6$$

| | | | | | | | |
|---|---|---|---|---|---|---|---|
| 0 | -0.16951 | | | | | | |
| 50 | -811.52448 | 9185.46392 | -43238.10254 | 108322.34424 | -152326.68058 | 114000.99795 | -35372.51408 |
| 100 | -1319.98652 | 14511.87509 | -66394.44524 | 161785.9296 | -221444.2-531 | 161422.90108 | -48957.00854 |
| 150 | -196.38457 | 2271.26657 | -10888.94575 | 27692.15437 | -39418.45425 | 29786.99366 | -9337.03939 |
| 200 | 54.22064 | -470.83555 | 1598.05375 | -2597.12854 | 1854.85502 | -166.56485 | -292.35386 |
| 250 | 250.05467 | -2585.63489 | 11105.7324 | -25372.54288 | 32511.47981 | -22149.49221 | 6267.65872 |
| 300 | 315.91956 | -3293.47776 | 14274.48684 | -32934.89604 | 42657.85719 | -29405.22298 | 8427.98342 |
| 350 | 684.91094 | -7256.77927 | 32002.81913 | -75207.55737 | 99328.54596 | -69904.23073 | 20481.4182 |
| 400 | 1027.94353 | -11011.86286 | 49107.27503 | -116702.13456 | 155870.44692 | -110934.7875 | 32868.84025 |
| 450 | 1160.57571 | -12352.82906 | 54723.27865 | -129162.21824 | 171300.50405 | -121033.53034 | 35593.16454 |
| 500 | 2473.34807 | -26601.80166 | 119108.44995 | -294188.39003 | 381080.59744 | -272299.0439 | 81000.14562 |
| 550 | 2172.15112 | -23270.53275 | 103776.63443 | -246601.53222 | 329310.33953 | -234313.496057 | 69400.26603 |
| 600 | 2168.37577 | -23128.45159 | 102683.62823 | -242898.30219 | 322869.35584 | -228653.20175 | 67400.89709 |
| 650 | 1759.28615 | -18684.24079 | 82600.96424 | -194578.33836 | 257578.61299 | -181675.75679 | 53338.88193 |
| 700 | 1347.72283 | -14180.3909 | 62097.58472 | -144876.38215 | 189915.36029 | -132625.63223 | 38546.92225 |
| 750 | 1313.08342 | -13818.48508 | 60519.67095 | -141198.73961 | 185078.73961 | -129220.86049 | 37544.11367 |
| 800 | 888.94482 | -9277.63502 | 40296.97192 | 93244.16415 | 121217.88814 | -83937.57509 | 24186.35437 |
| 850 | 550.03149 | -5666.5144 | 24293.14221 | -55483.39191 | 71190.05169 | -48651.20567 | 13834.64459 |
| 900 | 440.79781 | -4513.43272 | 19229.52192 | -43642.67912 | 55639.84347 | -37776.71793 | 10671.06066 |
| 950 | 320.85062 | -3293.51885 | 14077.62959 | -32078.13098 | 41086.70714 | -28041.50628 | 7966.44081 |
| 1000 | 198.6091 | -1994.94073 | 8341.64168 | -18593.5246 | 23292.91486 | -15545.82134 | 4318.14336 |